\documentclass[superscriptaddress,prl,twocolumn,showpacs,preprintnumbers,amsmath,amssymb]{revtex4}

\usepackage{graphicx}
\usepackage{color}

\begin{document}

\title{A topological method to characterize tapped granular media from the position of the particles.}

\author{S.Ardanza-Trevijano}\email{sardanza@unav.es}
\affiliation{Departamento de F\'{\i}sica, Facultad de Ciencias,
Universidad de Navarra, 31080 Pamplona, Spain.}

\author{Iker Zuriguel}\email{iker@unav.es}
\affiliation{Departamento de F\'{\i}sica, Facultad de Ciencias,
Universidad de Navarra, 31080 Pamplona, Spain.}

\author{Roberto Ar\'evalo}
\affiliation{CNR--SPIN, Dipartimento di Scienze Fisiche,
Universit\`a di Napoli Federico II, I-80126, Napoli, Italy
}

\author{Diego Maza}
\affiliation{Departamento de F\'{\i}sica, Facultad de Ciencias,
Universidad de Navarra, 31080 Pamplona, Spain.}

\date{\today}

\begin{abstract}

We use the first Betti number of a complex to characterize the
morphological structure of granular samples in mechanical
equilibrium. We analyze two-dimensional granular packings after a
tapping process by means of both simulations and experiments.
States with equal packing fraction obtained with different tapping
intensities are distinguished after the introduction of a
filtration parameter which determines the particles (nodes in the
network) that are joined by an edge. We first use numerical
simulations to characterize the effect of the precision in the
particles localization by artificially adding different levels of
noise in this magnitude. The outcomes obtained for the simulations
are then compared with the experimental results allowing a clear
distinction of experimental packings that have the same density.
This is accomplished by just using the position of the particles
and no other information about the possible contacts, or magnitude
of forces.
\end{abstract}


\maketitle

Any  dry granular material is a collection of macroscopic
particles interacting between them mainly by contact forces. Due
to this generality it is natural and appealing to consider a
granular system as a graph where contacts between particles are
edges, and the corresponding particles, nodes. Thus, defining such
a network can be useful as it can be analyzed using the machinery
of topology and modern complex networks theory \cite{newman}.
Based in the results of this analysis, the peculiarities and
generalities of a granular ensemble can be characterized.

The contact network was used by Adler \cite{adler} to study
transport in porous media, and by Dodds \cite{dodds} to analyze
the porosity in random sphere packings. More recently, Basset
\emph{et al.} \cite{daniels} use complex networks to understand
sound propagation in granular materials. The network of contacts
can also be used to characterize the evolution of granular media
in dynamic situations. Walker, Tordesillas and collaborators have
used simulations to deform granular samples under a variety of
conditions and complex networks tools to study the resulting
evolving networks \cite{walker1,walker2,tordesillas}. In
\cite{tordesillas1} the topological properties of the network were
related to the process of strain localization, which leads to
shear banding and material failure. Related with failure is the
process of buckling of force chains studied in
\cite{tordesillas2}, where it was revealed the importance of the
presence of loops of contacts in the network. These contact loops
were also proved to be crucial in the stability of tilted granular
samples \cite{smart}. Loops with an odd number of particles were,
indeed, previously proved to be important for the rigidity of
granular materials \cite{rivier}.

Additionally  to the contact network (the graph of all contacts),
the force network  can be also analyzed by the same methods.
Indeed,  the normalized contact force $f=F/<F>$ (where $F$ is the
force present in a contact and $<F>$ the sample average) can be
used to define as edge any contact bearing a force $f$ larger than
some threshold value $f^*$ that can be tuned in the range
$[0,f_{max}]$. Thus, for $f^*=0$ one recovers the contact network,
while for larger values one obtains progressively diluted graphs.
The analysis of the topology of the force networks has been shown
very fruitful. Ostojic \emph{et al.} \cite{ostojic} used it to
uncover a universality in the force distribution of granular
packing. A variety of topological measures as function of $f^*$
were analyzed in \cite{arevalo3} in a static granular packing. The
process of jamming in the light of the topology of force networks
was studied in \cite{arevalo4,arevalo1}. Again, the role of loops
in the network (particularly, third order loops) was proved to be
relevant at the transition point, with third order loops behaving
as an order parameter. Related to jamming is the question of
isostaticity, which was analyzed using the force network by Walker
\emph{et al.} \cite{tordesillas3}.

Recently, Kondic and coworkers have analyzed the force network
using topological invariants. In \cite{Kondic} the zeroth Betti
number $B0$, which measures the number of connected components
(clusters), is used to study compressed granular samples. $B0$ is
shown to be useful characterizing force networks obtained with
varying density, friction and polydispersity of the grains. The
zeroth Betti number is also used in \cite{kondic2} to analyze the
role of interparticle friction in impact dynamics. Carlsson
\emph{et al.} \cite{carlsson} used the zeroth and first Betti
numbers to analyze a $2D$ system with small number of particles.
Interestingly, they showed that critical points (where the
topology can change) correspond to configurations in mechanical
equilibrium.

It seems that studying contact and force networks offers a
fruitful pathway for the understanding of static and dynamic
properties of granular media. It is not completely clear, however,
if these tools provide more (or better) information than other
traditional measures. This question was addressed in
\cite{Arevalotappingtriangulos} where the topology
of $2D$ granular samples in mechanical equilibrium was studied. It was already
known \cite{pugnaloni1} that samples with the same density and
number of particles may not be in the same state of equilibrium since
the average force moment tensor can be different. In
\cite{Arevalotappingtriangulos} it was shown that the topology of
the contact network (without information on the forces) was enough
to distinguish these mechanically different states.
Interestingly, traditional measurements based on particles'
positions --like the pair correlation function, the bond order
parameter and the Voronoi tessellation-- were shown to be less
sensitive to capture such differences among different states with
the same packing fraction.

In the same line is the recent work of Kramar \emph{et al.}
\cite{KramarGoulletKondicMichaikow} who have used persistent
homology to study the evolution of the force network in compressed
granular materials. They track the evolution of the topology as a
function of the density for different polydispersities and
friction coefficients. Their approach is able to uncover the
distinctive behavior displayed by different systems and, moreover,
it is shown to be richer in information than the pair correlation
function, the bond orientational order parameter and the
distribution function of the forces.

Most of the works mentioned here are theoretical or consist on
numerical simulations. In these conditions one has all the
information necessary to construct the contact and force networks.
However, under experimental conditions it is difficult to
establish with certainty if there is contact between adjacent
particles. It is then desirable to devise a method to study the
contact network even in the case in which contacts cannot be exactly determined.
In the present work we aim at precisely
this goal using persistent homology.

Our system of study is a granular bed subject to tapping, which
has been widely studied; experimentally
\cite{nowak1,nowak2,richard,schroter}, by means of simulations
\cite{ciamarra,pugnaloni1,pugnaloni2}, and theoretically
\cite{edwards1}. In \cite{pugnaloni1,pugnaloni2} it was shown that
the packing fraction $\phi$ of the bed is not a monotonous
function of the tapping intensity $\Gamma$. This rises the
question of whether states with the same density are equivalent or
not. A negative answer to this question was given in
\cite{pugnaloni2} analyzing the force moment tensor of the system.
As mentioned before, the same result can be obtained using the
contact network. In the present work we use simulations and
experiments to show that even when the contacts among the
particles are not known, persistent homology allows to distinguish
between states with the same density but in different mechanical
equilibrium states.

\begin{center}
\begin{figure}
\includegraphics[scale=0.25]{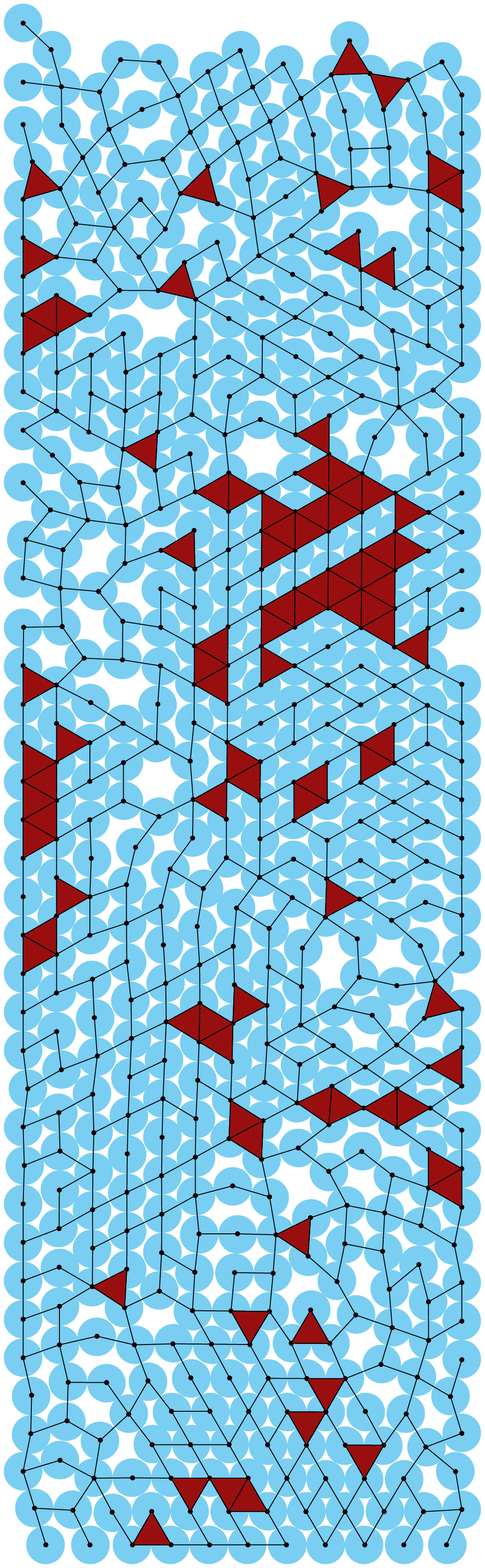}
\includegraphics[scale=0.25]{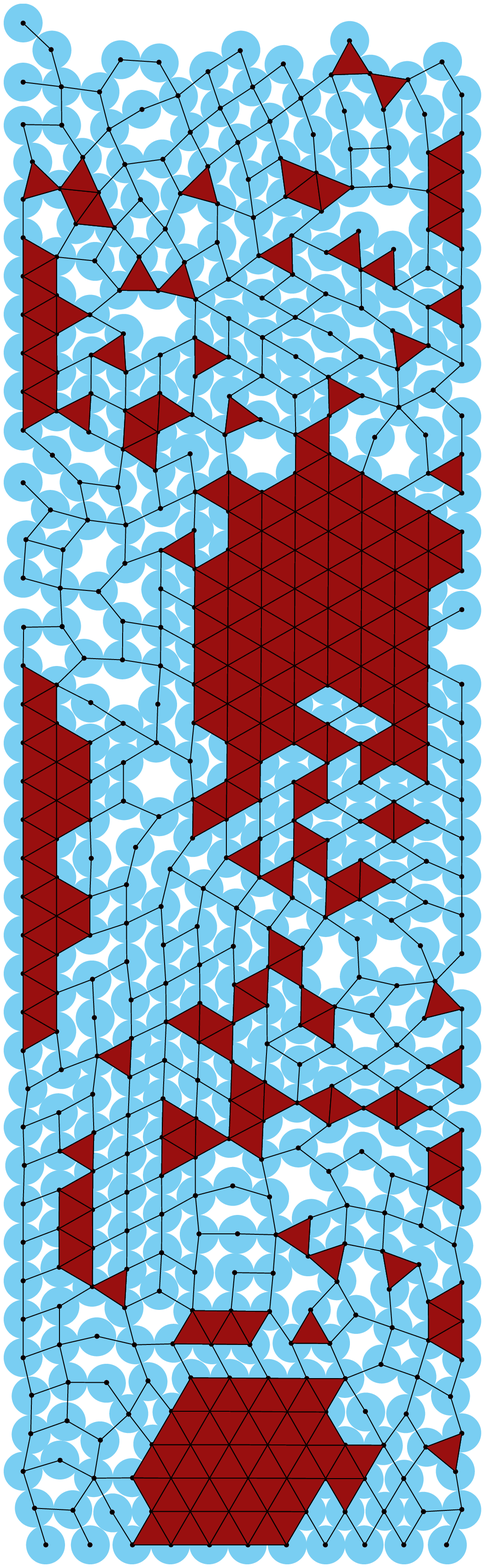}
\includegraphics[scale=0.25]{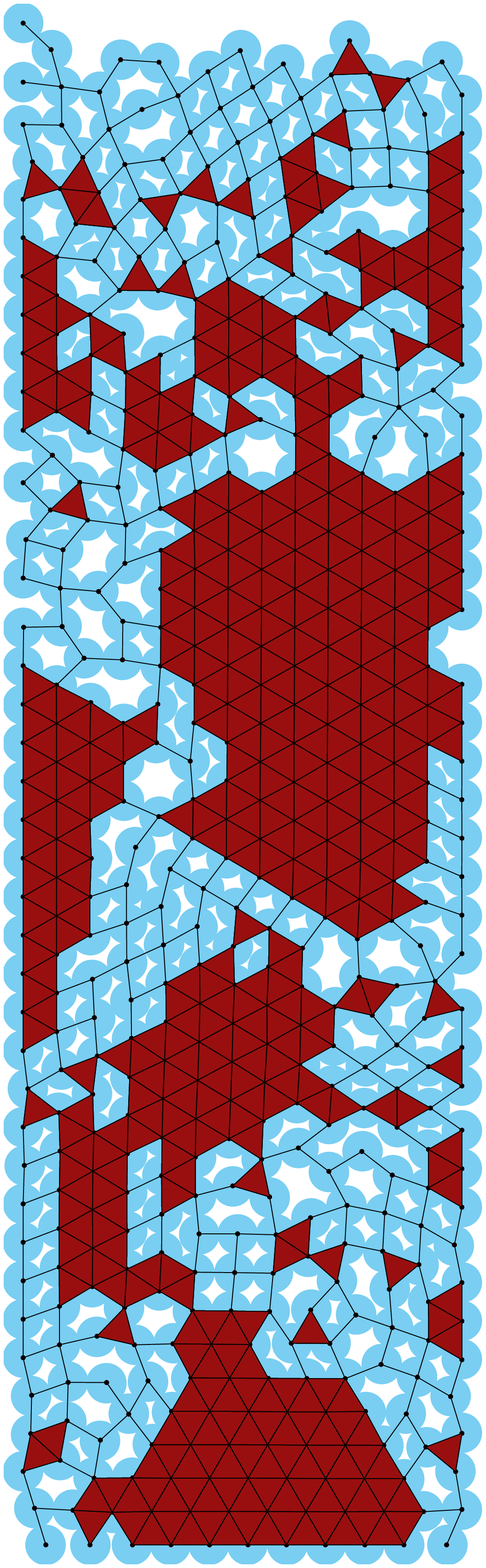}
\includegraphics[scale=0.24]{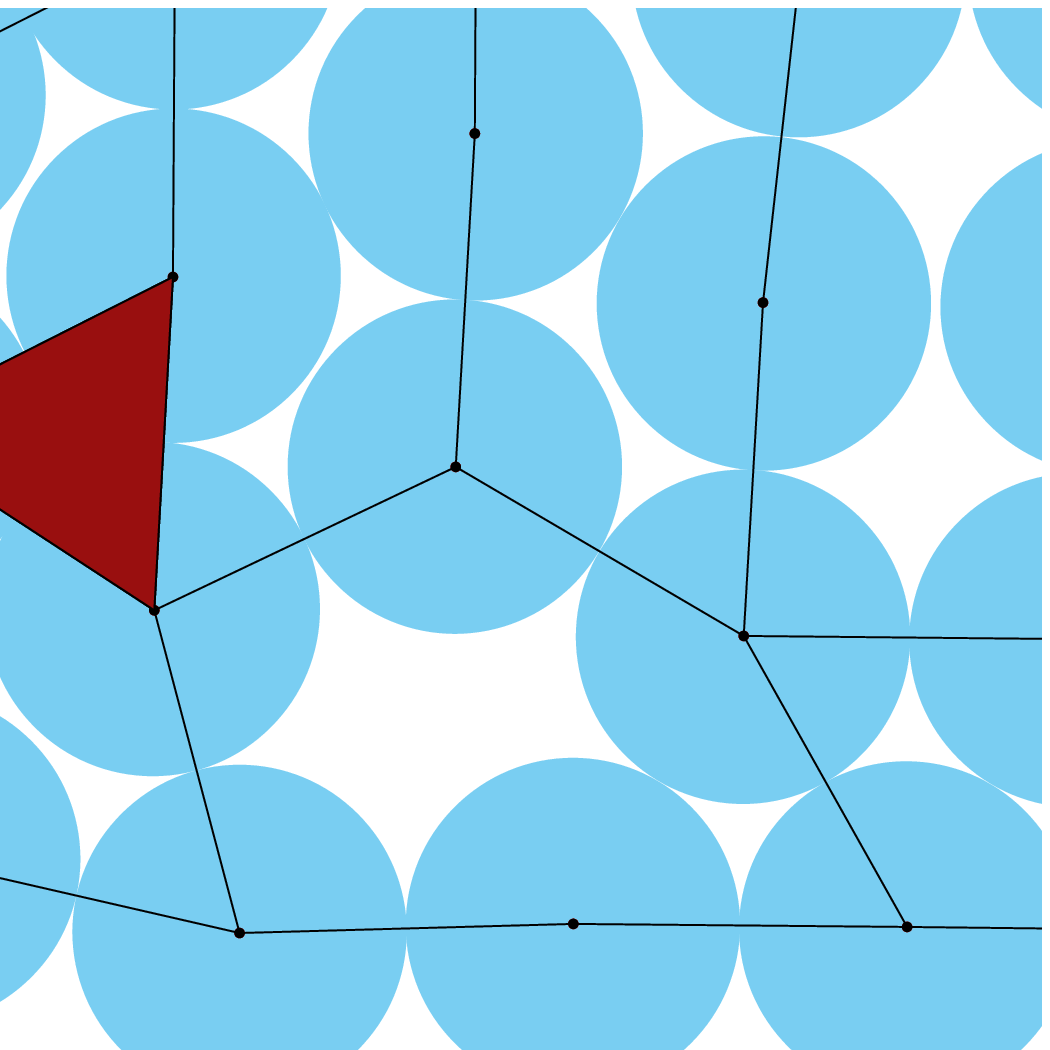}
\includegraphics[scale=0.24]{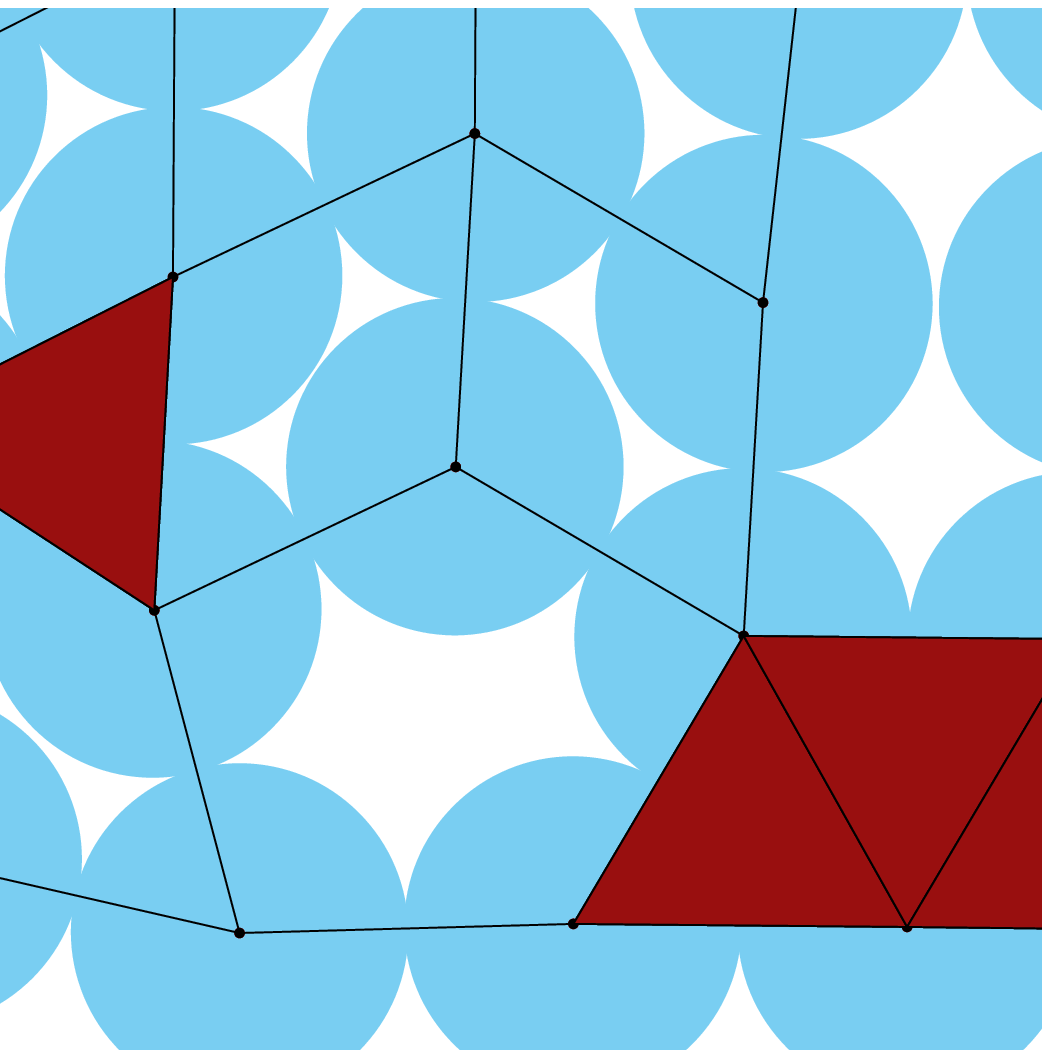}
\includegraphics[scale=0.24]{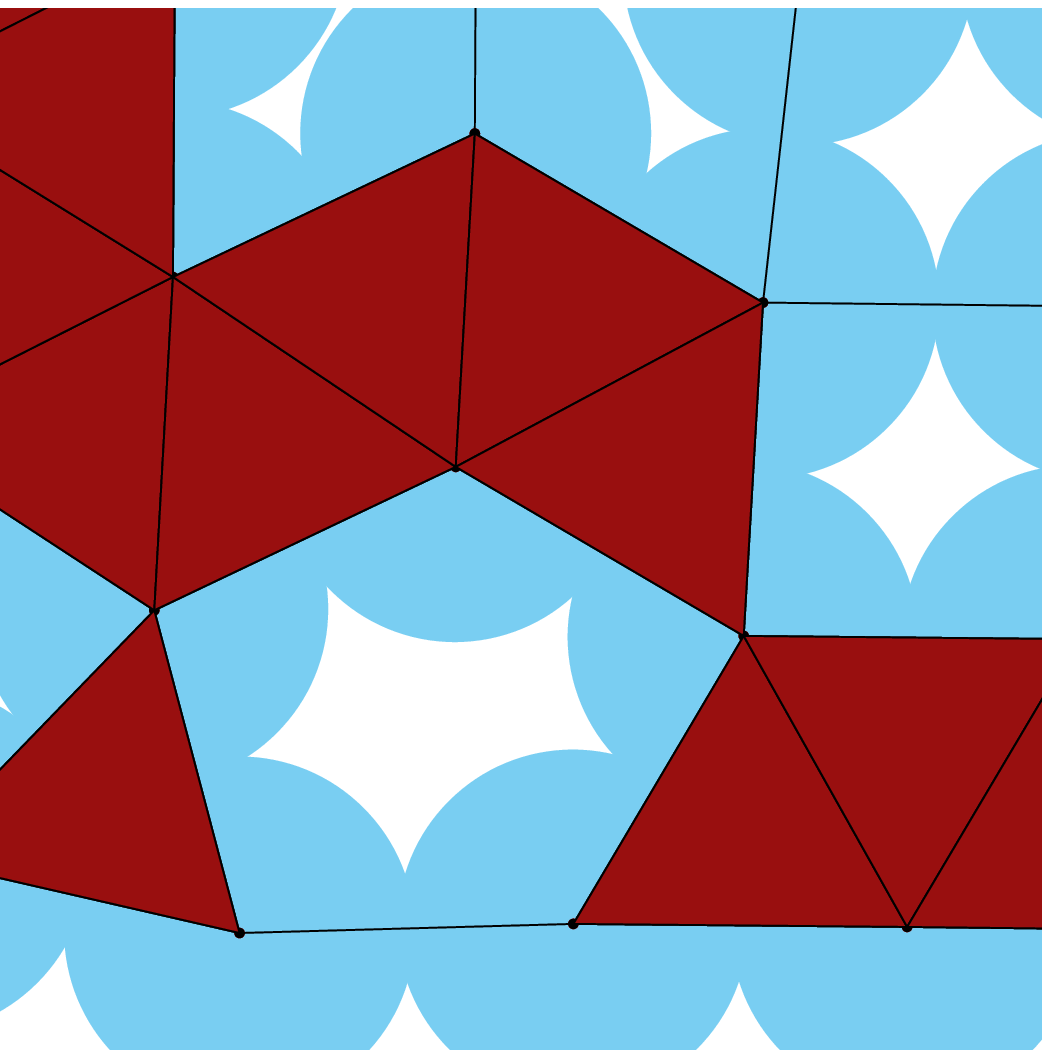}

\caption{(Color online) At the top: three examples of the
Vietoris-Rips complex (clique complex of the graph) obtained from
the data given by the centers of particles of a sample obtained
with $\Gamma=2$ which has an associated value of packing fraction
$\phi=0.8$. Each column corresponds to a different value of the
filtration parameter $\delta$. On the left $\delta=d$ the exact
diameter of the particle,  on the center $\delta=1.01 d$ and on
the right $\delta=1.05 d$. On the bottom, a magnified region is
shown evidencing the effect of increasing the filtration parameter
in the Vietoris-Rips complex: some quadrilaterals in the middle
picture are converted in triangles in the right picture. As
explained in the text, the creation of a triangle implies an
augment of $B1$ in the graph, and a decrease of $B1$ in the
associated clique complex. \label{Vietoris}}

\end{figure}
\end{center}

\section{Simulation and experimental methods}

\emph{Simulation protocol:} We use soft-particle molecular
dynamics simulations in $2D$, in which static friction is
implemented through the usual Cundall-Strack model \cite{cundall}.
The details of the implementation have been described elsewhere
\cite{arevalo2}, and in the following we give the values of the
interaction parameters used in the present work. In the normal
direction of the contact we set the stiffness $k_n=10^5(mg/d)$ and
damping parameter $\gamma_n=300(m\sqrt{g/d})$. In the tangential
direction we set $k_s=\frac{2}{7}k_n$ for the stiffness and
$\gamma_s=200(m\sqrt{g/d})$ for the damping parameter. The
friction coefficient is fixed to $\mu=0.5$. We used reduced units
with the diameter $d$ of the disks, the mass $m$ and the
acceleration of gravity $g$. The integration time step is
$\tau=10^{-4}\sqrt{d/g}$. The confining box has a width of
$13.39d$ and infinitely high lateral walls. Simulations are run
with $N=512$ monosized disks of diameter $d$.

The tapping is simulated by moving the confining box in the
vertical direction following a sine shaped trajectory $A
\sin(2\pi\nu t)[1-\Theta(2\pi\nu t-\pi)]$. We fix the frequency at
the value $\nu=\pi/2(g/d)^{1/2}$ and control the tapping intensity
$\Gamma=A(2\pi\nu)^2/g$ through the amplitude $A$. Once a tap is
applied we decide that the system is in equilibrium implementing a
criterium based on the stability of the contacts \cite{arevalo2}.
At this point, particle positions are recorded which will be
subsequently used to calculate both, the packing fraction and the
network properties. In particular, the packing fraction is
calculated in a slab of the bed that covers $50\%$ of the height
of the column and is centered on the center of mass of the system.
Then, a new tap is applied. Following this protocol we tap the bed
$1000$ times for each reported value of the intensity. Averages
are computed considering only the last $500$ taps of each run,
where (in all the cases) the packing fraction has already became
stationary, i.e. it has a well defined average.


In order to see how sensitive are our measurements to the lack of
precision in the determination of the particle centers, we use a
strategy consisting in adding controlled noise to the particles'
positions. From the original positions obtained from the
simulations, we created sets of noisy data in the following
manner. Given a value of $\alpha$ in $[0, 0.1]$, we moved each
center to a point at a random distance sampled from a uniform
distribution in $[0, \alpha\,d]$ where $d$ is the diameter of the
particles, and a random direction sampled from a uniform
distribution in $[0 ,2\pi]$.

\emph{Experimental setup:} A quasi 2D Plexiglass cell (width: 28
mm, height: 150 mm) was used to study the dependence of the
packing on the intensity of shaking. The cell was filled with 600
alumina oxide spheres of diameter $d=1.00\;mm$. The side wall
separation was 10\% larger than the bead diameter in order to
minimize the particle-wall friction and prevent arching in the
transversal direction. The system was tapped with an
electromagnetic shaker that provides a sine shaped pulse with a
frequency ($\nu$) and an amplitude ($A$). The frequency was kept
constant at ($\nu=30\;Hz$) and the amplitude was systematically
modified in order to vary the tapping intensity
$\Gamma=\frac{A(2\pi\nu)^{2}}{g}$. The latter was measured with a
piezoelectric accelerometer attached to the base of the cell. High
resolution digital images of the packings were taken after each
tap. The center of each sphere was detected with an error smaller
than $2\%$ of the particles diameter. The packing fraction of each
packing was calculated by considering each grain as a disk of the
corresponding effective diameter and then calculating the
percentage of the area covered by the disks in a rectangular area
10\% smaller than the size of the whole packing. In order to
determine the average packing fraction, for each tapping amplitude
we average over $200$ samples after reaching a stationary packing
fraction. More details of the experimental protocol can be found
in \cite{pugnaloni2}.

\section{Persistent Homology}

Persistent homology is a tool that provides topological
information of an object examined at different resolutions. We
will give an ad-hoc description in the following paragraph and
recommend the interested reader the sources
\cite{Zomorodian,EdelsbrunnerHarer,Cannon} for a more detailed and
broad description. Since our data are 2D we will restrict all the
relevant constructions to two dimensions.

Our data is the position of the centers of the particles in an experiment or simulation i.e. a set of points in the plane.
The natural way to  build a contact network is to consider the graph that has as vertices the mentioned set of points, and add an edge
between a pair of vertices $p_i$, $p_j$, if the euclidean distance between them is less than the diameter $d$ of the particles $d(p_i,p_j)\leq d$. However this construction will miss some existing contacts and add some non existing contacts due to noise and/or numerical precision. To deal with this problem we construct a parametrized collection of graphs where the vertices are the same in all graphs (the above mentioned set of points),  but edges in each graph are added whenever the distance between two points is less than a parameter $\delta\geq0$.

Three particles that are in contact with each other form a ``local
perfect packing''. In order to keep track of these, we build a
second structure associated with each one of the previously
described graphs. If three nodes (particles) have all pairwise
connections, i.e. edges between them form a triangle, we add a
2D-cell covering the triangle. We thus obtain a sort of
``tesselation with holes'' (see Fig. \ref{Vietoris}), called the
clique complex of the graph in which the ``holes'' are the
polygons formed by closed loops in the graph that are not
triangles. This structure is a 2D-simplicial complex that is
usually called the 2D-clique complex of the graph, and also the
2D-Vietoris-Rips complex of the dataset for the given filtration
parameter.

%

Once we have a simplicial complex we can calculate its Betti
numbers which are non negative integers, one for each dimension.
Since our complexes live in the Euclidean plane, we are only
interested in 0 dimensional and 1 dimensional Betti numbers.  The
0-th Betti number ($B0$) of a complex is the number of connected
components, and the 1-st Betti number ($B1$) counts the number of
1D-holes (the network polygons) in our complex. We will calculate
the first Betti number (B1) of both the graph and  the clique
complex. In the graph, this accounts for the total number of
1D-holes, i.e. the number of polygons given by edges connecting
data points in the graph. In the clique complex, B1 provides the
number of uncovered polygons, i.e. polygons that are not
triangles. Due to the presence of gravity, we expect to have a
single connected component in most cases ($B0=1$), and thus we focus
our study only in $B1$.

The effect of increasing the filtration parameter above the
diameter of the particles in the $B1$ of the graph is that the
development of new connections necessarily leads to the apparition
of polygons and hence, to the increase of $B1$. In the clique
complex, however, new connections may lead to the creation of
polygons, but also to the covering of a triangle (and hence to a
reduction of $B1$). We study filtration parameters in the range
$d\leq \delta \leq 1.1d$ where $d$ denotes the diameter of the
particles both, in the simulations and in the experiments. In this
article, the homology calculations have been performed using the
software \emph{javaplex} \cite{javaplex}, developed by the group
of Applied and Computational Algebraic Topology of Stanford
University.

\section{Simulation results}

\begin{figure}
\includegraphics[scale=0.6]{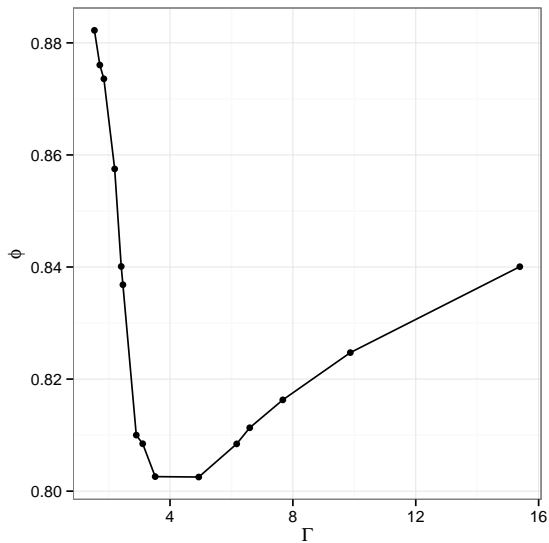}
\caption{Mean packing fraction $\phi$ of the steady states as a
function of the tap intensity $\Gamma$ obtained from numerical
simulations. \label{phivsgamma_simul}}
\end{figure}

Let us start presenting in Fig. \ref{phivsgamma_simul} already
known results of the packing fraction $\phi$ dependence on the
tapping intensity $\Gamma$ \cite{pugnaloni1}. There is a non
monotonous dependence of $\phi$, which first decreases with
$\Gamma$ and then, increases after a given value of the excitation
parameter, $\Gamma_{min}$. This implies that states with the same
packing fraction are obtained with very different tapping
intensities. These states of equal $\phi$ have been demonstrated
to display completely different stress properties
\cite{pugnaloni2}, being also distinguishable by measuring the
number of polygons of $3$, $4$, $5$, $6$... sides, obtained from
the contact network \cite{Arevalotappingtriangulos}.

\begin{figure}
\includegraphics[scale=0.6]{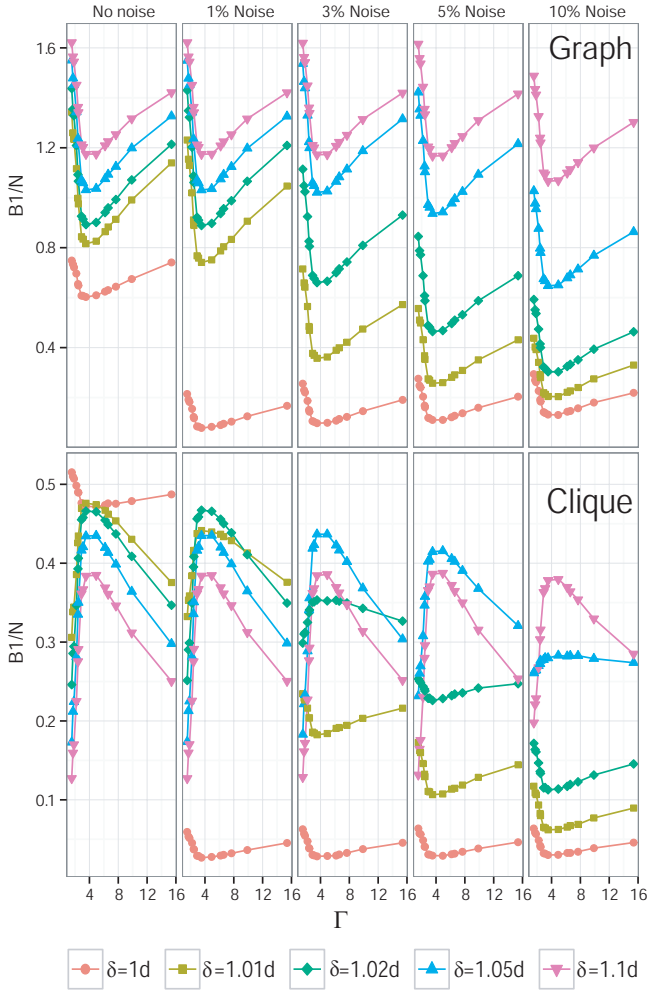}
\caption{(Color online) Simulation results of the mean first Betti
number (B1) normalized by the number of particles as a function of
the tap intensity ($\Gamma$) for different values of the
filtration parameter $\delta$ and different levels of noise in the
data. At the top, results obtained from the graph and at the
bottom, results obtained from the clique. The 95\% confidence
intervals for the mean of the normalized Betti numbers are of the
size of the data points. \label{betti1vsgamma_simul}}
\end{figure}

In what follows we explain how persistent homology is useful to
unveil topological differences among these states without the
necessity of having an exact identification of all the contacts in
the network. In Fig. \ref{betti1vsgamma_simul} results of mean B1
(normalized by the number of particles) are presented versus
$\Gamma$ for different values of the filtration parameter $\delta$
(as indicated in the legend) and different levels of noise
(increasing from left to right panels). Looking at the results of
the graph obtained for $\delta=1d$ without noise (circles in the
top left panel), we realize the same qualitative behavior than the
obtained for the total number of polygons in the contact network
\cite{Arevalotappingtriangulos}. At the same panel, we observe
that increasing the value of $\delta$ leads to an augment of the
values of B1 (the number of polygons increases) preserving the
shape of the curves. Furthermore, when noise is added to the data
the curve trends of $B1$ versus $\Gamma$ in the graph are
maintained (top figures of Fig. \ref{betti1vsgamma_simul}).
Importantly, the only effect that increasing the level of noise
seems to have in the $B1$ graph is a downwards displacement of the
curves (thus, the effect of random noise is to destroy polygons of
all types). Adding $1\%$ noise strongly (slightly) affects
$\delta=1.00d$ ($\delta=1.01d$) curves and has not apparent effect
on curves obtained for higher values of $\delta$. Adding $3\%$
noise strongly affects $\delta=1.00d$, $\delta=1.01d$, and
$\delta=1.02d$ curves; adding $5\%$ noise strongly affects curves
with $\delta<1.05d$, and weakly affects $\delta=1.05d$. Despite
the analysis of the noise effect may appear rather artificial at
this point, it will be proved to be relevant when analyzing the
experimental data which are, intrinsically, noisy.

A rather different behavior is obtained when displaying the values
of B1 for the clique (bottom panels in Fig.
\ref{betti1vsgamma_simul}). We will start explaining the case
without noise (bottom left panel). Interestingly, although the
trend displayed for $\delta=1d$ is similar to the obtained in the
graph, a small increase of $\delta$ leads to a change of the curve
trend: the minimum is transformed into a maximum. Considering that
the only difference between the graph and the clique is that the
latter does not account for triangles, the comparison of the
correspondent curves provides interesting information. Focusing in
the case of $\delta=1.01d$, the fact that the B1 of the clique
increases with $\Gamma$ and then, after $\Gamma_{min}$ decreases
again, implies that the number of polygons --without considering
triangles- is maximum in $\Gamma_{min}$. At this same point, the
total number of polygons (B1 of the graph) was proved to be
minimum. This reflects that, the increase in B1 of the graph
obtained when we move apart from $\Gamma_{min}$ is due to an
augment in the number of triangles, and a consequent reduction in
the number of the other polygons. A further increase of $\delta$
(which leads to increasing values of B1 in the graph) provokes a
reduction of B1 in the clique without alteration of the curve
trend. This evidences that most of the polygons that are build in
the graph when increasing $\delta$ are, indeed, triangles.

\begin{figure}
\includegraphics[scale=0.6]{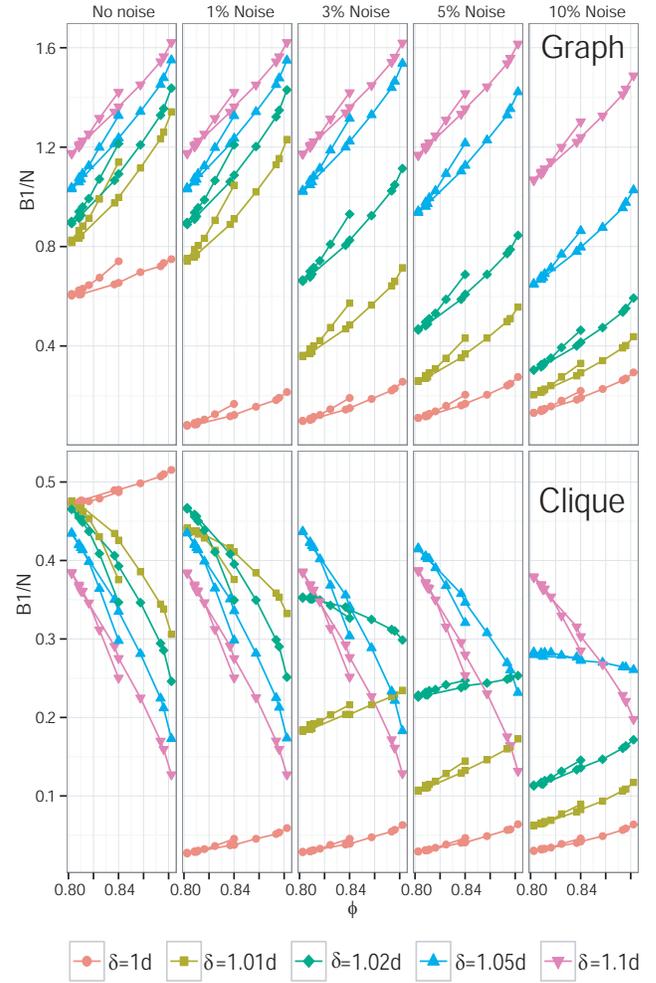}
\caption{(Color online) Simulation results of the mean first Betti
number (B1) normalized by the number of particles as a function of
the packing fraction ($\phi$) for different values of the
filtration parameter $\delta$ and different levels of noise in the
data. At the top, results obtained from the graph and at the
bottom, results obtained from the clique. The 95\% confidence
intervals for the mean of the normalized Betti numbers are of the
size of the data points. \label{phivsbetti1_simul}}
\end{figure}

In the clique curves, the effect of adding noise is also notably
different from that observed in the graph. If the value of
$\delta$ is higher than the level of noise, the curves show a
maximum and the values of B1 are reduced as $\delta$ increases. On
the contrary, if $\delta$ is smaller than the noise level, the
curves that originally displayed a maximum invert their shape and
show a minimum -- revealing a trend similar to the one observed
for the case without noise and $\delta=1.00d$. This effect can be
explained as follows. First, it should be recalled that for the
case without noise, increasing $\delta$ leads to the development
of a maximum in the clique curves as a consequence of the increase
in the number of triangles. Considering this, it seems reasonable
that adding a given amount of noise destroys some of the triangles
creating polygons of any kind. The only way to compensate the
addition of noise (and preserve the triangular structure in the
network) is applying a  sufficiently high filtration parameter.

\begin{figure}
\includegraphics[scale=0.6]{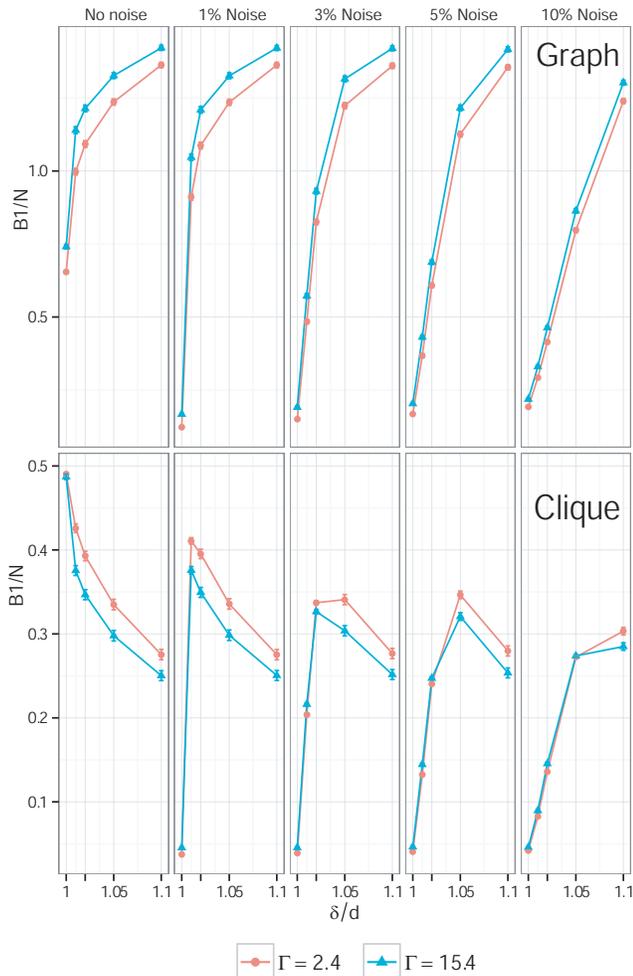}
\caption{(Color online) Comparison of the mean first Betti number
normalized by the number of particles for states with the same
packing fraction obtained with different tap intensities
($\Gamma=2.4$ and $\Gamma=15.4$). Results obtained from
simulations are presented versus the value of the filtration
parameter for different levels of noise as indicated at the top of
each figure. At the top, results obtained from the graph and at
the bottom, results obtained from the clique. The error bars
indicate the 95\% confidence intervals for the mean of the
normalized Betti numbers.
\label{bettisimulcomparebettivsdelta2gammas}}
\end{figure}

\begin{figure*}[t]
\includegraphics[scale=1.2]{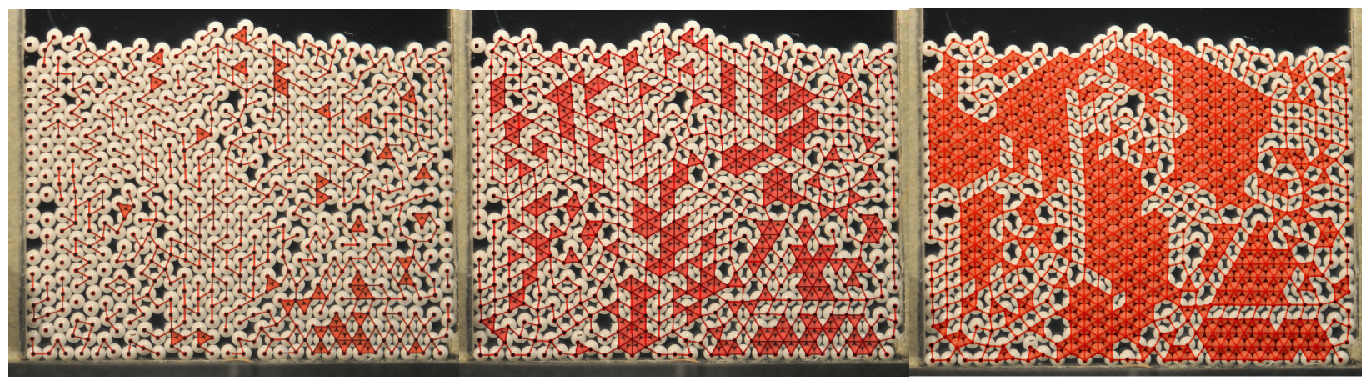}
\caption{(Color online) Experimental photograph with the
associated Vietoris-Rips complex for three different values of the
filtration parameter. Left $\delta=d$, the estimated diameter of
the particles, center $\delta=1.01 d$ and right $\delta=1.05 d$.
\label{fotoexp}}
\end{figure*}

Since the objective of this manuscript is the implementation of
persistent homology to distinguish among states with the same
packing fraction, it seems more practical representing B1 (for
different values of noise and filtration parameter) with respect
to the packing fraction $\phi$ (Fig. \ref{phivsbetti1_simul}).
Interestingly, all the B1 data present two well defined  branches; in this case
the shorter one is for high $\Gamma$ and the longer one for low
$\Gamma$ (see Fig. \ref{phivsgamma_simul} to realize that the
values of $\phi$ obtained above $\Gamma_{min}$ span over a smaller
region than the ones obtained below $\Gamma_{min}$). These
branches are more or less separated from each other depending on
the values of noise and filtration parameter. Focusing first on
the results of the graph without noise for $\delta=1.00d$, we
observe that B1 increases with $\phi$, but this increment is more
pronounced for the short branch (higher values of $\Gamma$).
Comparing these results with the analogous of the clique, where
the two branches are undistinguishable, we can conclude that the
differences among the two branches are predominantly caused by the
development of triangles (which are more abundant for high values
of $\Gamma$). This result agrees whith the analysis
carried out in \cite{Arevalotappingtriangulos}.

The effect of increasing $\delta$ in the graph obtained without
noise is just an augment of the values of B1 without changing the
shape of the curves. Nevertheless, an exceedingly high filtration
parameter like $\delta=1.10d$ seems to provoke a reduction in the
separation between the two branches (down triangles in the top
left graph of Fig. \ref{phivsbetti1_simul}). The introduction of
noise induces a decrease of the B1 values that mainly affects the
curves obtained with a filtration parameter smaller than the level
of noise. Notably, the decrease of B1 within a given curve is
rather homogeneous, being independent on $\phi$. This effect was
already appreciated in Fig. \ref{betti1vsgamma_simul}.

In the data obtained from the clique, the effect of adding noise
and changing the filtration parameter may lead to an inversion of
the tendency of the curves. Focusing first in the the case without
noise, if $\delta>1.00d$ B1 decreases with $\phi$ in contrast to
the case of $\delta=1.00d$. The origin of this change (which was
already explained when describing the results displayed Fig
\ref{betti1vsgamma_simul}) is based in the development of
triangles for high values of $\phi$ attained for very high and
very low values of $\Gamma$. The introduction of noise in the
system leads to the transition from ascendent to descendent curves
appearing for larger values of $\delta$. More interestingly, it
seems that given a value of noise, the differences among the two
branches in the clique networks are maximized for a filtration
parameter higher or similar to the level of noise.

In order to check this idea, let us compare the outcomes of the B1
for two states that, being obtained with very different tap
intensities, display the same packing fraction. This occurs, for
example, for the states developed for $\Gamma=2.4$ and
$\Gamma=15.4$ whose B1 values for different noise and filtration
parameters are presented in Fig.
\ref{bettisimulcomparebettivsdelta2gammas}. Clearly, the results
obtained for the graph (top figures) reveal differences, being the
values of B1 systematically higher for the highest tapping
intensity. However, the differences become more or less important
depending on the noise and the filtration parameter. For the data
without additional noise, it seems that the outcomes of B1 are
already different for $\delta=1d$. The differences are magnified
for larger values of $\delta$ and seem to be minimized again for
$\delta=1.1d$. Similar trends are observed when noise is added in
the data. For these cases, however, it should be emphasized that
the differences for $\delta=1d$ become almost nonexistent. Indeed,
as the levels of noise augment, distinguishing among the states
requires of larger values of $\delta$.

The results of the clique (bottom of Fig.
\ref{bettisimulcomparebettivsdelta2gammas}) reveal that, opposite
to the graph, the B1 values are systematically smaller for the
case of the highest tap intensity. Again, this reveals that states
with the same packing fraction develop more triangles when
obtained at high tap intensities. Concerning the differences among
the states when adding noise and changing the filtration
parameter, the conclusions attained from the clique are similar to
those already explained for the graph. In summary, for low levels
of noise, differences are maximized for intermediate values of
$\delta$. As the noise is increased, the values of $\delta$ from
which differences appear also increase. The curve trends
(monotonously decreasing for the case without added noise, and
displaying a maximum when some noise is added) can be explained as
a consequence of the development of triangles in the network.
Indeed, the fact that for the case without noise, the B1 of the
clique decreases with $\delta$ while B1 of the graph increases, is
just a result of the development of triangles attained when
increasing $\delta$. When adding noise, a drastic reduction is
obtained in the B1 values of the clique for $\delta=1d$. This
reduction was already observed in the graph, and thus it is due to
the destruction of most of the polygons in the network. Of course,
the values of $\delta$ which suffer modifications in the B1
outcomes, are only those smaller or similar to the level of noise.
This leads to the non monotonicity of the curves.

From the numerical simulations we have learnt that the values of
B1 obtained from both, the graph and the clique, for different
levels of noise and filtration parameter, can be very useful to
characterize the topology of the systems. Indeed, systems with the
same packing fraction obtained with different intensities are
clearly differentiated as they present different branches in the
B1 versus $\phi$ plot. The important advance of the present work
is that we can dodge the exact determination of all the contacts
in the network. Remarkably, we have shown that, even in the
presence of an important amount of noise in the particles'
positions, we can identify differences in the systems by
appropriately defining a filtration parameter.

\section{Experimental results}

Once we have seen that states with the same packing fraction but
obtained from very different tapping intensities can be
distinguished even in the presence of considerable amount of
noise, we turn on experimental data, which have a precision in the
determination of the centers of the particles smaller than $2\%$
of their diameter. In Fig. \ref{fotoexp} we show an experimental
photograph with the Vietoris-Rips complex obtained using three
values of the filtration parameter $\delta$. The one on the left
uses $\delta=d$ leading to a quite unrealistic network due to the
lack of precision in the determination of the particles' centers.
As $\delta$ increases, more contacts appear in the complex.
Obviously, some of them are spurious as they are not real
contacts, specially for the right picture where $\delta=1.05 d$ is
employed. These three pictures evidence the difficulty of properly
defining a network of contacts from experimental data, suggesting
the necessity of looking for alternatives that help to
characterize the structure of the packing.

\begin{figure}
\includegraphics[scale=0.6]{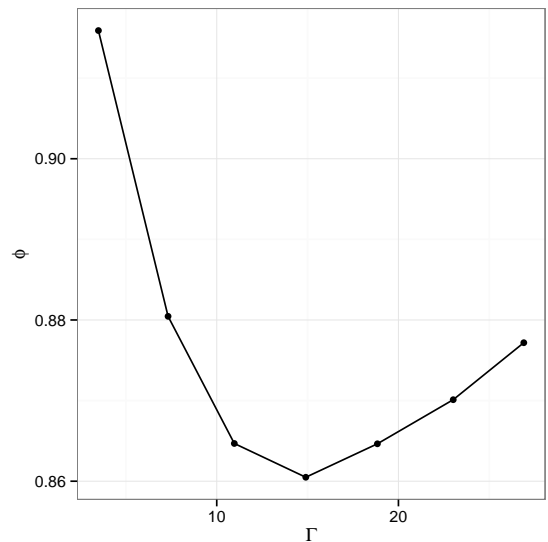}
\caption{Experimental results of the mean packing fraction $\phi$
of the steady states as a function of the tap intensity $\Gamma$.
\label{phivsgamma_exp600}}
\end{figure}

\begin{figure}[t]
\includegraphics[scale=0.6]{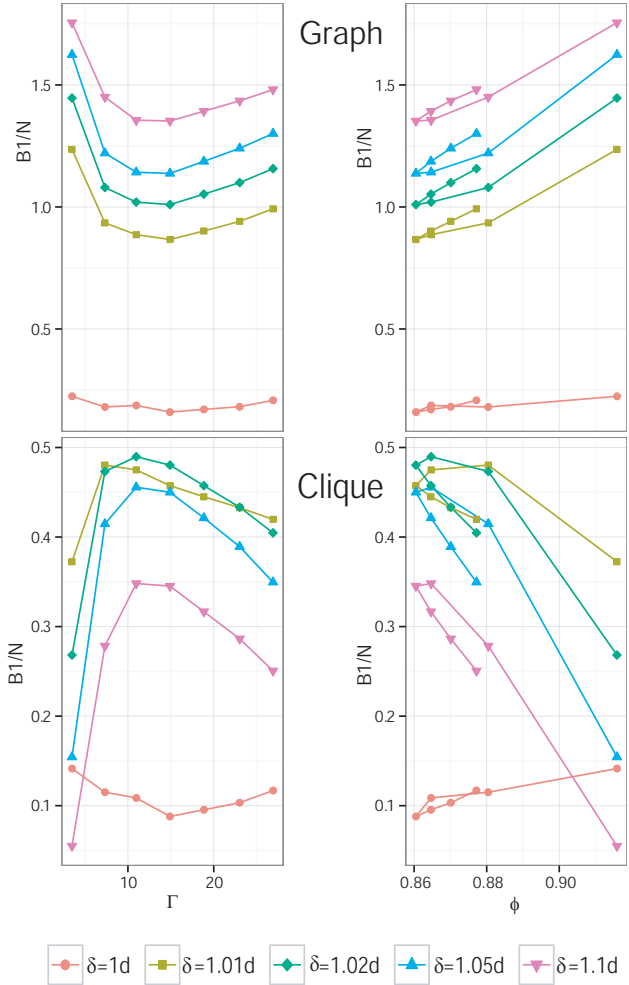}
\caption{(Color online) Experimental results of the mean first
Betti number (B1) normalized by the number of particles for
different values of the filtration parameter $\delta$ as indicated
in the legend. In the left column, B1 is presented versus the tap
intensity $\Gamma$. In the right column B1 is presented versus the
packing fraction of the sample $\phi$. At the top, results
obtained from the graph and at the bottom, results obtained from
the clique. In all cases, the 95\% confidence intervals for the
mean of the normalized Betti numbers are of the size of the data
points. \label{betti1vsgammaandfi_exp}}
\end{figure}

\begin{figure}
\includegraphics[scale=0.6]{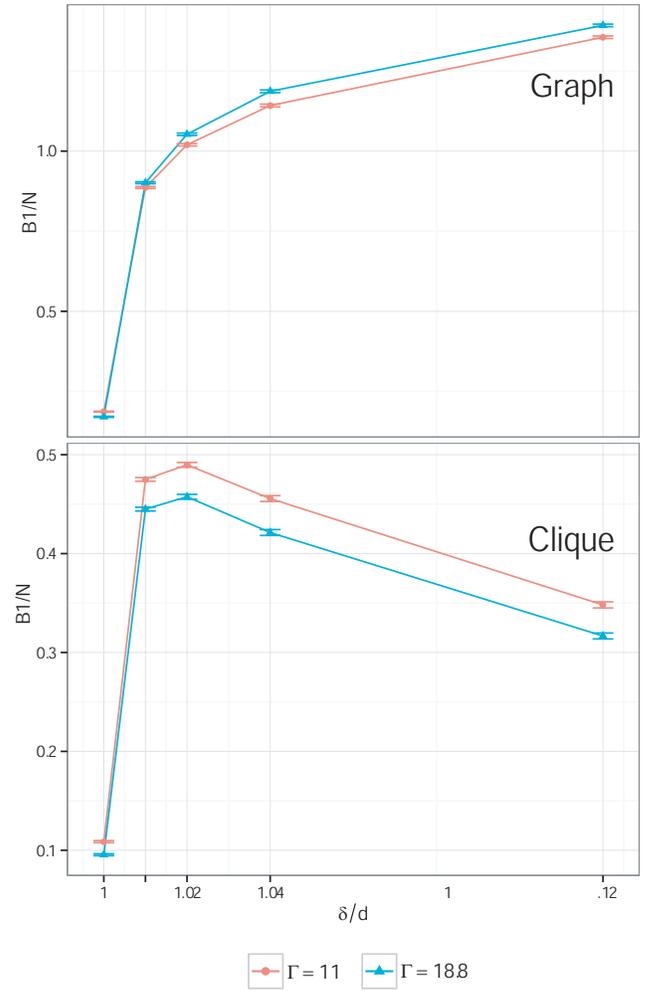}
\caption{(Color online) Mean first Betti number versus $\delta$
for states with the same average packing fraction obtained with
different tap intensities. As indicated in the legend, squares
(triangles) are used for the state reached with $\Gamma=11$
($\Gamma=18.8$). At the top, results obtained from the graph and
at the bottom, results obtained from the clique. The error bars
indicate the 95\% confidence intervals for the mean of the
normalized Betti numbers.
\label{bettiexpcomparebettivsdelta2gammas}}
\end{figure}

In Fig. \ref{phivsgamma_exp600} we show experimental results of
the average packing fraction of the steady state as a function of
$\Gamma$. As observed in the simulations, the packing fraction
first decreases with $\Gamma$, and then increases after
$\Gamma_{min}\approx15$. Hence, states with the same $\phi$ are
obtained with very different tapping intensities. The goal now is
testing if these states can be distinguished by means of the
average first Betti number as explained above. The evolution of B1
versus the tapping intensity is presented in the left column of
Fig. \ref{betti1vsgammaandfi_exp} for both, the graph (top) and
the clique (bottom). The outcomes strikingly resemble the
numerical results obtained when adding $1\%$ noise to the
particle's position (second column in Fig.
\ref{betti1vsgamma_simul}). As in the simulations, in the graph,
the B1 curve obtained for $\delta=1.00 d$ is significatively below
the curves obtained for higher $\delta$. In addition, in the
clique, the curve for $\delta=1.00 d$ shows a minimum at
$\Gamma_{min}$ that is converted into a maximum for $\delta=1.01
d$. As explained in the previous section, this maximum is
consequence of the presence of triangles in the network which is
specially important for very high and low values of $\Gamma$
(those that produce large values of $\phi$).

The accordance of the experimental data with the numerical results
displayed for the case of $1\%$ noise is confirmed by the plots of
B1 versus $\phi$ (right column of Fig.
\ref{betti1vsgammaandfi_exp}). Surely, the graph shows an increase
of B1 with $\phi$ independently on the value of $\delta$, but the
two developed branches are clearly distinguished for values of the filtration
parameter $\delta>1.00 d$. In the clique, for $\delta>1.00 d$, B1
decreases with $\phi$ as a consequence of the presence of
triangles as explained previously. Again, increasing the value of
$\delta$ we can achieve a clear differentiation among the two
branches obtained in the B1 versus $\phi$ plot.

Finally, in Fig. \ref{bettiexpcomparebettivsdelta2gammas} we
compare the experimental values of B1 for two states with the same
packing fraction but obtained at different tap intensities, i.e.
$\Gamma=11$ and $\Gamma=18.8$ for the left and right sides of
$\Gamma_{min}$. In the graph, the B1 values obtained for the
highest $\Gamma$ are systematically above those obtained for the
lowest $\Gamma$. This trend is reversed for the clique, where the
B1 values obtained for the highest $\Gamma$ are systematically
below those obtained for the lowest $\Gamma$. This behavior is in
perfect agreement with numerical simulations, and can be
attributed to the number of triangles developed in the network
which is more important for the higher value of $\Gamma$.
Interestingly, it is observed that a good election of the
filtration parameter is crucial in order to differentiate among
states with the same packing fraction. Once more, the experimental
results are compatible with the outcomes of the simulations with a
noise of approximately $1\%$. This implies that values of
$\delta>1.01 d$ are necessary to observe an enhancement of the
differences among the two states.

\section{Conclusions}

In this work, we have shown that the first Betti number of the
graph and the clique (the Vietoris-Rips complex) can be
satisfactorily used to characterize granular packings. Using a
filtration parameter that defines whether or not two particles in
the sample (nodes) are joined by a link, we are able to
differentiate among states that display the same packing fraction
but which are, indeed, different. This is accomplished even when
the contacts among the particles are not accesible due to the lack
of precision in the determination of the particles' positions. In
the first part of the manuscript we have studied the B1 dependence
on both $\Gamma$ and $\phi$ by means of numerical simulations
where the particle's position can be obtained with $10^{-8}d$
precision. Then, we have artificially introduced noise in the
positions of the particles and characterized its effect in the
observed behavior. Based on these results we have implemented the
same techniques on experimental data, finding qualitative
agreement with the numerical outcomes for the case of $1\%$ noise.

More importantly, the results reported in this work prove that an
accurate determination of the contacts among the particles is not
necessary to observe topological differences among states with the
same packing fraction, but obtained with different tapping
intensities. This result supposes an important step forward with
respect to a previous one \cite{Arevalotappingtriangulos} where
the topological tool introduced to identify such differences is
only available if the contact network is well defined. On the
contrary, the topological approach introduced in this manuscript
can be used to characterize experimental packing ensembles even
where the contact network is not fully accessible due to the
limited resolution of the experimental techniques.

\begin{acknowledgments}
RA thanks MIUR-FIRB RBFR081IUK for financial support. This work
has been financially supported by Projects FIS2011-26675 (Spanish
Government) and PIUNA (Universidad de Navarra).
\end{acknowledgments}

\end{document}